\documentstyle[12pt]{article}
\begin{document}

\author{Yishi Duan\thanks{%
E-mail: ysduan@lzu.edu.cn} and Ying Jiang\thanks{%
Corresponding author, E-mail: itp3@lzu.edu.cn} \\
{\small {\it Institute of Theoretical Physics, Lanzhou University, Lanzhou
730000, P. R. China}}}
\title{The generation of the $(k-1)$--dimensional defect objects and their
topological quantization\thanks{%
This work is supported by the National Natural Science Foundation of P. R.
China} }
\date{}
\maketitle

\begin{center}
\begin{minipage}{150mm}
\noindent
\_\hrulefill\\
\noindent
{\small
{\bf Abstract}

\baselineskip20pt
In the light of $\phi $--mapping method and
topological current theory, the topological structure and the topological
quantization of arbitrary dimensional topological defects are investigated.
It is pointed out that the topological quantum numbers of
the defects are described by the Winding numbers of $\phi $--mapping which
are determined in terms of the Hopf indices and the Brouwer degrees of
$\phi$--mapping.
Furthermore, it is shown that all the topological defects are
generated from where $\vec \phi =0$, i.e. from the zero points of the $\phi $%
--mapping.

\vskip3mm
\noindent
{\it PACS}: 11.27 +d; 02.40 -k; 04.20 -q; 98.80 Cq\\
{\it Keywords}: Topological defects; $\phi$--mapping; topological current; topological quantization}

\noindent
\_\hrulefill
\end{minipage}
\end{center}

\vskip0.5cm \baselineskip24pt

\section{Introduction}

The world of topological defects is amazingly rich and have been the focus
of much attention in many areas of contemporary physics\cite{cos,cosd,gue}.
The importance of the role of defects in understanding a variety of problems
in physics is clear\cite{jan,wang,blamo,cen}. So, it is necessary for us to
investigate the topological properties of the topological defects
meticulously. Recently, some physicists noticed\cite{zyg,neil} that the
topological defects are closely related to the spontaneously broken of $O(m)$
symmetry group to $O(m-1)$ by $m$--component order parameter field $\vec
\phi $ and pointed out that for $m=1$, one has domain walls, $m=2$, strings
and $m=3$, monopoles, for $m=4$, there are textures. But for the lack of a
powerful method, the topological properties are not very clear yet.

In this paper, in the light of $\phi $--mapping topological current theory%
\cite{11}, a useful method which plays a important role in studying the
topological invariants\cite{lee,li} and the topological structures of
physical systems\cite{14,zhh,4}, we will investigate the topological
quantization and the evolution of these topological defects.

\section{The generalized topological current}

As is well known, in our previous papers, only the topological current of
point-like particles was discussed. In this paper, we will extend the
concept to present an arbitrary dimensional generalized topological current.
We consider the $\phi$--mapping as a map between two manifolds, while the
dimensions of the two manifolds are arbitrary. It is an important
generalization of our previous work on topological current and is of great
usefulness to theoretical physics and differential geometry.

In $n$--dimensional Riemann manifold $G$ with metric tensor $g_{\mu \nu }$
and local coordinates $x^\mu $ $(\mu ,\nu =1,...,n)$, a $m$--component
vector order parameter $\vec \phi (x)$ can be looked upon as a mapping
between the Riemann manifold $G$ and a $m$--dimensional Euclidean space $R^m$%
$$
\phi :\;G\rightarrow R^m,\;\;\;\;\;\;\phi ^a=\phi ^a(x),\;\;\;a=1,...,m. 
$$
The direction field of $\vec \phi (x)$ is generally determined by 
\begin{equation}
\label{0}n^a(x)=\frac{\phi ^a(x)}{||\phi (x)||},\;\;\;\;\;||\phi (x)||=\sqrt{%
\phi ^a(x)\phi ^a(x)} 
\end{equation}
with 
\begin{equation}
\label{1}n^a(x)n^a(x)=1. 
\end{equation}
It is obviously that $n^a(x)$ is a section of the sphere bundle $S(G)\cite
{11}$. If $n^a(x)$ is a smooth unit vector field without singularities or it
has singularities somewhere but at the point $\vec \phi (x)\neq 0$, from (%
\ref{1}) we have 
\begin{equation}
\label{2}n^a\partial _\mu n^a=0,\;\;\;\;\mu =1,...,n, 
\end{equation}
which can be looked upon as a system of $n$ homogeneous linear equations of $%
n^a$ $(a=1,...,m)$ with coefficient matrix $[\partial _\mu n^a]$. The
necessary and sufficient condition that (\ref{2}) has non--trivial solution
for $n^a(x)$ is rank $[\partial _\mu n^a]<m$, i.e. the Jacobian determinants 
\begin{equation}
\label{3}D^{\mu _1\cdot \cdot \cdot \mu _k}(\partial n)=\frac 1{m!}\epsilon
^{\mu _1\cdot \cdot \cdot \mu _k\mu _{k+1}\cdot \cdot \cdot \mu _n}\epsilon
_{a_1\cdot \cdot \cdot a_m}\partial _{\mu _{k+1}}n^{a_1}\cdot \cdot \cdot
\partial _{\mu _n}n^{a_m} 
\end{equation}
are equal to zero, where $k=n-m$. while, at the point $\vec \phi =0$, the
above consequences are not held. In short, we have the following relations 
\begin{equation}
\label{4}D^{\mu _1\cdot \cdot \cdot \mu _k}(\partial n)\left\{ 
\begin{array}{cc}
=0, & \;for\; \vec \phi \neq 0, \\ 
\neq 0, & \;for\;\vec \phi =0, 
\end{array}
\right. 
\end{equation}
which implies $D^{\mu _1\cdot \cdot \cdot \mu _k}(\partial n)$ behaves
itself like a function $\delta (\vec \phi )$. So we are focussed on the
zeroes of $\phi ^a(x)$. Suppose that the vector field $\vec \phi (x)$
possesses $l$ zeroes, according to the implicit function theorem\cite{golsat}%
, when the zeroes are regular points of $\phi $--mapping at which the rank
of the Jacobian matrix $[\partial _\mu \phi ^a]$ is $m$, the solutions of $%
\vec \phi =0$ can be expressed parameterizedly by 
\begin{equation}
\label{5}x^\mu =z_i^\mu (u^1,\cdot \cdot \cdot ,u^k),\;\;\;\;i=1,...,l, 
\end{equation}
where the subscript $i$ represents the $i$--th solution and the parameters $%
u^I$ ($I=1,...,k$) span a $k$--dimensional submanifold with the metric
tensor $g_{IJ}=g_{\mu \nu }\frac{\partial x^\mu }{\partial u^I}\frac{%
\partial x^\nu }{\partial u^J}$ which is called the $i$--th singular
submanifold $N_i$ in the Riemannian manifold $G$ corresponding to the $\phi $%
--mapping. For each singular manifold $N_i$, we can define a normal
submanifold $M_i$ in $G$ which is spanned by the parameters $v^A$ with the
metric tensor $g_{AB}=g_{\mu \nu }\frac{\partial x^\mu }{\partial v^A}\frac{%
\partial x^\nu }{\partial v^B}$ $(A,B=1,...,m)$, and the intersection point
of $M_i$ and $N_i$ is denoted by $p_i$. In fact, in the words of
differential topology, $M_i$ is transversal to $N_i$ at the point $p_i$. By
virtue of the implicit function theorem, at the regular point $p_i$, it
should be hold true that the Jacobian matrix $J(\frac \phi v)$ satisfies 
\begin{equation}
\label{nonzero}J(\frac \phi v)=\frac{D(\phi ^1,\cdot \cdot \cdot ,\phi ^m)}{%
D(v^1,\cdot \cdot \cdot ,v^m)}\neq 0. 
\end{equation}

In the following, we will induce a rank--$k$ topological current through the
integration of $D^{\mu _1\cdot \cdot \cdot \mu _k}(\partial n)$ in (\ref{3})
on $M_i$. As is well known, the generalized Winding Number\cite{19} has been
given by the Gauss map $n:\partial \Sigma _i\rightarrow S^{m-1}$%
\begin{equation}
\label{6}W_i=\frac 1{A(S^{m-1})(m-1)!}\int_{\partial \Sigma
_i}n^{*}(\epsilon _{a_1\cdot \cdot \cdot a_m}n^{a_1}dn^{a_2}\wedge \cdot
\cdot \cdot \wedge dn^{a_m}) 
\end{equation}
where%
$$
A(S^{m-1})=\frac{2\pi ^{m/2}}{\Gamma (m/2)} 
$$
is the area of ($m-1$)--dimensional unit sphere $S^{m-1}$, $n^{*}$ denotes
the pull back of map $n$ and $\partial \Sigma _i$ the boundary of a
neighborhood $\Sigma _i$ of $p_i$ on $M_i$ with $p_i\notin \partial \Sigma
_i $, $\Sigma _i\cap \Sigma _j=\emptyset $. The generalized Winding Numbers $%
W_i $ can also be rewritten as%
$$
W_i=\frac 1{A(S^{m-1})(m-1)!}\int_{n[\partial \Sigma _i]}\epsilon _{a_1\cdot
\cdot \cdot a_m}n^{a_1}dn^{a_2}\wedge \cdot \cdot \cdot \wedge dn^{a_m} 
$$
which means that, when the point $x^\mu $ or $v^A$ covers $\partial \Sigma
_i $ once, the unit vector $n^a$ will cover a region $n[\partial \Sigma _i]$
whose area is $W_i$ times of $A(S^{m-1})$, i.e. the unit vector $n^a$ will
cover the unit sphere $S^{m-1}$ $W_i$ times. From the above equation, one
can deduce that%
\begin{eqnarray}
W_i&=&\frac 1{A(S^{m-1})(m-1)!}\int_{\partial M_i}\epsilon _{a_1\cdot \cdot
\cdot a_m}n^{a_1}\partial _{\mu _{k+2}}n^{a_2}\cdot \cdot \cdot \partial
_{\mu _n}n^{a_m}dx^{\mu _{k+2}}\wedge \cdot \cdot \cdot \wedge dx^{\mu _n}\nonumber \\
&=&\frac 1{A(S^{m-1})(m-1)!}\int_{M_i}\frac 1{k!}\frac 1{\sqrt{g_x}}\epsilon
^{\mu _1\cdot \cdot \cdot \mu _k\mu _{k+1}\cdot \cdot \cdot \mu _n}\epsilon
_{a_1\cdot \cdot \cdot a_m}\partial _{\mu _{k+1}}n^{a_1}\partial _{\mu
_{k+2}}n^{a_2}\cdot \cdot \cdot \partial _{\mu _n}n^{a_m}d\sigma _{\mu
_1\cdot \cdot \cdot \mu _k}\nonumber \\
\label{7}&=&\frac 1{A(S^{m-1})(m-1)!}\int_{M_i}\frac 1{k!}\frac{m!}{\sqrt{g_x}}%
D^{\mu _1\cdot \cdot \cdot \mu _k}(\partial n)d\sigma _{\mu _1\cdot \cdot
\cdot \mu _k}, 
\end{eqnarray}
where $d\sigma _{\mu _1\cdot \cdot \cdot \mu _k}$ is the invariant surface
element of $M_i$ and $g_x=\det (g_{\mu \nu })$. As mentioned above, the
deduction (\ref{7}) shows that the Winding Numbers $W_i$ can be expressed as
the integration of $D^{\mu _1\cdot \cdot \cdot \mu _k}(\partial n)$ on $M_i$.

From the above discussions, especially the expressions (\ref{3}), (\ref{4})
and (\ref{7}), we can induce a generalized topological current $j^{\mu
_1\cdot \cdot \cdot \mu _k}$ which does not vanish only at the zeroes of the
order parameter field $\vec \phi (x)$, and is exactly corresponding to the
generalized Winding Number, 
\begin{equation}
\label{8}j^{\mu _1\cdot \cdot \cdot \mu _k}=\frac 1{A(S^{m-1})(m-1)!\sqrt{g_x%
}}\epsilon ^{\mu _1\cdot \cdot \cdot \mu _k\mu _{k+1}\cdot \cdot \cdot \mu
_n}\epsilon _{a_1\cdot \cdot \cdot a_m}\partial _{\mu _{k+1}}n^{a_1}\partial
_{\mu _{k+2}}n^{a_2}\cdot \cdot \cdot \partial _{\mu _n}n^{a_m}. 
\end{equation}
Obviously this tensor current is identically conserved, i.e.%
$$
\nabla _{\mu _i}j^{\mu _1\cdot \cdot \cdot \mu _k}=0,\;\;\;i=1,...,k. 
$$
The dual tensor of $j^{\mu _1\cdot \cdot \cdot \mu _k}$ is 
$$
\tilde j_{\nu _1\cdot \cdot \cdot \nu _m}=\frac 1{A(S^{m-1})(m-1)!}\epsilon
_{a_1\cdot \cdot \cdot a_m}\partial _{\nu _1}n^{a_1}\partial _{\nu
_2}n^{a_2}\cdot \cdot \cdot \partial _{\nu _m}n^{a_m} 
$$
with 
\begin{equation}
\label{9}j^{\mu _1\cdot \cdot \cdot \mu _k}=\frac 1{\sqrt{g_x}}\epsilon
^{\mu _1\cdot \cdot \cdot \mu _k\nu _1\cdot \cdot \cdot \nu _m}\tilde j_{\nu
_1\cdot \cdot \cdot \nu _m}. 
\end{equation}
It is easy to see that both $j^{\mu _1\cdot \cdot \cdot \mu _k}$ and $\tilde
j_{\nu _1\cdot \cdot \cdot \nu _m}$ are completely antisymmetric tensors.

\section{Topological quantization of defect objects}

By making use of the $\phi $--mapping theory, we will study the global
property of the generalized topological current $j^{\mu _1\cdot \cdot \cdot
\mu _k}$ on the whole manifold $G$ and conclude that $j^{\mu _1\cdot \cdot
\cdot \mu _k}$ behaves itself like the generalized function $\delta (\vec
\phi )$ and the integration of $\tilde j$ is the Winding Numbers at
singularities $z(u)$ of $n^a(x)$. From (\ref{0}) we have%
$$
\partial _\mu n^a=\frac 1{||\phi ||}\partial _\mu \phi ^a+\phi ^a\partial
_\mu (\frac 1{||\phi ||}),\;\;\;\frac \partial {\partial \phi ^a}(\frac
1{||\phi ||})=-\frac{\phi ^a}{||\phi ||^3} 
$$
which should be looked upon as generalized functions\cite{gelfand}. Using
these expressions the generalized topological current (\ref{8}) can be
rewritten as%
\begin{eqnarray}
j^{\mu _1\cdot \cdot \cdot \mu _k}&=&C_m\frac 1{\sqrt{g_x}}\epsilon ^{\mu
_1\cdot \cdot \cdot \mu _k\mu _{k+1}\cdot \cdot \cdot \mu _n}\epsilon
_{a_1\cdot \cdot \cdot a_m}\nonumber \\
& &\cdot \partial _{\mu _{k+1}}\phi ^a\partial _{\mu _{k+2}}\phi ^{a_2}\cdot \cdot
\cdot \partial _{\mu _n}\phi ^{a_m}\frac \partial {\partial \phi ^a}\frac 
\partial {\partial \phi ^{a_1}}(G_m(||\phi ||)),\;\;\;\;\;m>2.\nonumber
\end{eqnarray}
where $C_m$ is a constant%
$$
C_m=\left\{ 
\begin{array}{cc}
-\frac 1{A(S^{m-1})(m-2)(m-1)!}, & \;\;\;\;m>2 \\ 
\frac 1{2\pi }, & \;\;\;\;m=2 
\end{array}
,\right. 
$$
and $G_m(||\phi ||)$ is a generalized function%
$$
G_m(||\phi ||)=\left\{ 
\begin{array}{ccc}
\frac 1{||\phi ||^{m-2}} & ,\;\;\; & m>2 \\ 
\ln ||\phi || & ,\;\;\; & m=2 
\end{array}
.\right. 
$$
Defining general Jacobians $J^{\mu _1\cdot \cdot \cdot \mu _k}(\frac \phi x)$
as following%
$$
\epsilon ^{a_1\cdot \cdot \cdot a_m}J^{\mu _1\cdot \cdot \cdot \mu _k}(\frac
\phi x)=\epsilon ^{\mu _1\cdot \cdot \cdot \mu _k\mu _{k+1}\cdot \cdot \cdot
\mu _n}\partial _{\mu _{k+1}}\phi ^{a_1}\partial _{\mu _{k+2}}\phi
^{a_2}\cdot \cdot \cdot \partial _{\mu _n}\phi ^{a_m} 
$$
and by making use of the $m$--dimensional Laplacian Green function relation%
\cite{11}%
$$
\Delta _\phi (\frac 1{||\phi ||^{m-2}})=-\frac{4\pi ^{m/2}}{\Gamma (\frac
m2-1)}\delta (\vec \phi ) 
$$
where $\Delta _\phi =(\frac{\partial ^2}{\partial \phi ^a\partial \phi ^a})$
is the $m$--dimensional Laplacian operator in $\phi $--space, we do obtain
the $\delta $--function like topological current rigorously 
\begin{equation}
\label{10}j^{\mu _1\cdot \cdot \cdot \mu _k}=\frac 1{\sqrt{g_x}}\delta (\vec
\phi )J^{\mu _1\cdot \cdot \cdot \mu _k}(\frac \phi x). 
\end{equation}
We find that $j^{\mu _1\cdot \cdot \cdot \mu _k}\neq 0$ only when $\vec \phi
=0$ ( or when $x\in N_i$), which is just the singularity of $j^{\mu _1\cdot
\cdot \cdot \mu _k}$. In detail, the Kernel of the $\phi $--mapping is the
singularities of the topological tensor current $j^{\mu _1\cdot \cdot \cdot
\mu _k}$ in $G$. We think that this is the essential of the topological
tensor current theory and $\phi $--mapping is the key to study this theory.

As is well known\cite{17}, the definition of the $\delta $--function $\delta
(N_i)$ in curved space-time on a submanifold $N_i$ is 
\begin{equation}
\label{m}\delta (N_i)=\int_{N_i}\frac 1{\sqrt{g_x}}\delta ^n(\vec x-\vec
z_i(u^1,u^2))\sqrt{g_u}d^ku,\;\;\;\;g_u=\det (g_{IJ}). 
\end{equation}
Following this, by analogy with the procedure of deducing $\delta (f(x))$,
since 
\begin{equation}
\delta (\vec \phi )=\left\{ 
\begin{array}{cc}
+\infty , & for\; \vec \phi (x)=0 \\ 
0, & for\;\vec \phi (x)\neq 0 
\end{array}
\right. =\left\{ 
\begin{array}{cc}
+\infty , & for\;x\in N_i \\ 
0, & for\;x\notin N_i 
\end{array}
\right. , 
\end{equation}
we can expand the $\delta $--function $\delta (\vec \phi )$ as 
\begin{equation}
\label{delta}\delta (\vec \phi )=\sum_{i=1}^lc_i\delta (N_i), 
\end{equation}
where the coefficients $c_i$ must be positive, i.e. $c_i=\mid c_i\mid $.
From the definition of $W_i$ in (\ref{6}), the Winding number can also be
rewritten in terms of the parameters $v^A$ of $M_i$ as 
$$
W_i=\frac 1{2\pi }\int_{\Sigma _i}\epsilon ^{A_1\cdot \cdot \cdot
A_m}\epsilon _{a_1\cdot \cdot \cdot a_m}\partial _{A_1}n^{a_1}\cdot \cdot
\cdot \partial _{A_m}n^{a_m}d^mv, 
$$
Then, by duplicating the above process, we have 
\begin{equation}
\label{W}W_i=\int_{\Sigma _i}\delta (\vec \phi )J(\frac \phi v)d^mv. 
\end{equation}
Substituting (\ref{delta}) into (\ref{W}), and considering that only one $%
p_i\in \Sigma _i$, we can get 
\begin{equation}
W_i=\int_{\Sigma _i}c_i\delta (N_i)J(\frac \phi v)d^mv=\int_{\Sigma
_i}\int_{N_i}c_i\frac 1{\sqrt{g_x}\sqrt{g_v}}\delta ^n(\vec x-\vec
z_i(u^1,u^2))J(\frac \phi v)\sqrt{g_u}d^ku\sqrt{g_v}d^mv. 
\end{equation}
where $g_v=\det (g_{AB})$. Because $\sqrt{g_u}\sqrt{g_v}d^kud^mv$ is the
invariant volume element of the Product manifold $M_i\times N_i$, so it can
be rewritten as $\sqrt{g_x}d^nx$. Thus, by calculating the integral and with
positivity of $c_i$, we get 
\begin{equation}
c_i=\frac{\beta _i\sqrt{g_v}}{\mid J(\frac \phi v)_{p_i}\mid }=\frac{\beta
_i\eta _i\sqrt{g_v}}{J(\frac \phi v)_{p_i}}, 
\end{equation}
where $\beta _i=|W_i|$ is a positive integer called the Hopf index\cite{20}
of $\phi $-mapping on $M_i,$ it means that when the point $v$ covers the
neighborhood of the zero point $p_i$ once, the function $\vec \phi $ covers
the corresponding region in $\vec \phi $-space $\beta _i$ times, and $\eta
_i=signJ(\frac \phi v)_{p_i}=\pm 1$ is the Brouwer degree of $\phi $-mapping%
\cite{20}. Substituting this expression of $c_i$ and (\ref{delta}) into (\ref
{10}), we gain the total expansion of the rank--$k$ topological tensor
current 
\begin{equation}
j^{\mu _1\cdot \cdot \cdot \mu _k}=\frac 1{\sqrt{g_x}}\sum_{i=1}^l\frac{%
\beta _i\eta _i\sqrt{g_v}}{J(\frac \phi v)|_{p_i}}\delta (N_i)J^{\mu _1\cdot
\cdot \cdot \mu _k}(\frac \phi x). 
\end{equation}
From the above equation, we conclude that the inner structure of $j^{\mu
_1\cdot \cdot \cdot \mu _k}$ is labelled by the total expansion of $\delta
(\vec \phi )$, which includes the topological information $\beta _i$ and $%
\eta _i.$

It is obvious that, in (\ref{5}), when $u^1$ and$\,u^I(I=2,...,k)$ are taken
to be time-like evolution parameter and space-like parameters, respectively,
the inner structure of $j^{\mu _1\cdot \cdot \cdot \mu _k}$ just represents $%
l$ $(k-1)$--dimensional topological defects moving in the $n$--dimensional
Riemann manifold $G$. The $k$-dimensional singular submanifolds $%
N_i\,\,(i=1,\cdot \cdot \cdot l)$ are their world sheets. Here we see that
the defect objects are generated from where $\vec \phi =0$ and, the Hopf
indices $\beta _i$ and Brouwer degree $\eta _i$ classify these defects. In
detail, the Hopf indices $\beta _i$ characterize the absolute values of the
topological quantization and the Brouwer degrees $\eta _i=+1$ correspond to
defects while $\eta _i=-1$ to antidefects. It must be pointed that the
relationship between the zero points of the $m$--dimensional order parameter
field $\vec \phi $ and the space location of these topological defects is
distinct and clear and it is obtained rigorously without tie on any concrete
model or hypothesis.

\section{Evolution equation of the defect objects}

At the beginning of this section, we firstly give some useful relations to
study many defects theory. On the $i$-th singular submanifold $N_i$ we have%
$$
\phi ^a(x)|_{N_i}=\phi ^a(z_i^1(u),\cdot \cdot \cdot z_i^n(u))\equiv 0, 
$$
which leads to%
$$
\partial _\mu \phi ^a\frac{\partial x^\mu }{\partial u^I}|_{N_i}=0,\;\,\;\;%
\;\;I=1,...,k 
$$
Using this relation and the expression of the Jacobian matrix $J(\frac \phi
v)$, we can obtain 
$$
J^{\mu _1\cdot \cdot \cdot \mu _k}(\frac \phi x)|_{\vec \phi =0}=\frac
1{m!}\epsilon ^{\mu _1\cdot \cdot \cdot \mu _k\mu _{k+1}\cdot \cdot \cdot
\mu _n}\epsilon _{a_1\cdot \cdot \cdot a_m}\frac{\partial \phi ^{a_1}}{%
\partial x^{\mu _{k+1}}}\cdot \cdot \cdot \frac{\partial \phi ^{a_m}}{%
\partial x^{\mu _n}} 
$$
$$
=\frac 1{m!}\epsilon ^{\mu _1\cdot \cdot \cdot \mu _k\mu _{k+1}\cdot \cdot
\cdot \mu _n}\epsilon _{a_1\cdot \cdot \cdot a_m}\frac{\partial \phi ^{a_1}}{%
\partial v^{A_1}}\cdot \cdot \cdot \frac{\partial \phi ^{a_m}}{\partial
v^{A_m}}\frac{\partial v^{A_1}}{\partial x^{\mu _{k+1}}}\cdot \cdot \cdot 
\frac{\partial v^{A_1}}{\partial x^{\mu _n}} 
$$
\begin{equation}
\label{35}=\frac 1{m!}\epsilon ^{\mu _1\cdot \cdot \cdot \mu _k\mu
_{k+1}\cdot \cdot \cdot \mu _n}\epsilon _{A_1\cdot \cdot \cdot A_m}J(\frac
\phi v)\frac{\partial v^{A_1}}{\partial x^{\mu _{k+1}}}\cdot \cdot \cdot 
\frac{\partial v^{A_1}}{\partial x^{\mu _n}}, 
\end{equation}
then from this expression, the rank--$k$ tensor current can be expressed by 
\begin{equation}
\label{36}j^{\mu _1\cdot \cdot \cdot \mu _k}=\frac 1{m!\sqrt{g_x}%
}\sum_{i=1}^l\beta _i\eta _i\sqrt{g_v}\delta (N_i)\epsilon ^{\mu _1\cdot
\cdot \cdot \mu _k\mu _{k+1}\cdot \cdot \cdot \mu _n}\epsilon _{A_1\cdot
\cdot \cdot A_m}\frac{\partial v^{A_1}}{\partial x^{\mu _{k+1}}}\cdot \cdot
\cdot \frac{\partial v^{A_1}}{\partial x^{\mu _n}}. 
\end{equation}
Thus, using the above formulas and (\ref{9}), on $M_i$, the integral of the
dual tensor $\tilde j_{\nu _1\cdots \nu _m}$ gives the following result%
$$
\int_{M_i}\tilde j_{\nu _1\cdots \nu _m}dx^{\nu _1}\wedge \cdots \wedge
dx^{\nu _m}=\beta _i\eta _i. 
$$
This shows that, for the first time, we gain the topological charges of
these defects which are determined by the Hopf indices and Brouwer degrees
of the $\phi $--mapping.

Corresponding to the rank--$k$ topological tensor currents $j^{\mu _1\cdot
\cdot \cdot \mu _k}$, it is easy to see that the Lagrangian of many defects
is just%
$$
L=\sqrt{\frac 1{k!}g_{\mu _1\nu _1}\cdot \cdot \cdot g_{\mu _k\nu _k}j^{\mu
_1\cdot \cdot \cdot \mu _k}j^{\nu _1\cdot \cdot \cdot \nu _k}}=\delta (\vec
\phi ) 
$$
which includes the total information of arbitrary dimensional topological
defects in $G$ and is the generalization of Nielsen's Lagrangian\cite{22}.
The action in $G$ is expressed by 
$$
S=\int_GL\sqrt{g_x}d^nx=\sum_{i=1}^l\beta _i\eta _i\int_{N_i}\sqrt{g_u}%
d^ku=\sum_{i=1}^l\beta _i\eta _iS_i 
$$
where $S_i$ is the area of the singular manifold $N_i$. It must be pointed
out here that the Nambu--Goto action\cite{13}, which is the basis of many
works on defect theory, is derived naturally from our theory. From the
principle of least action, we obtain the evolution equations of many defect
objects 
\begin{equation}
\label{38}g^{IJ}\frac{\partial g_{\nu \lambda }}{\partial x^\mu }\frac{%
\partial x^\nu }{\partial u^I}\frac{\partial x^\lambda }{\partial u^J}%
-2\frac 1{\sqrt{g_u}}\frac \partial {\partial u^I}(\sqrt{g_u}g^{IJ}g_{\mu
\nu }\frac{\partial x^\nu }{\partial u^J})=0,\;\;\;I,J=1,...,k. 
\end{equation}
As a matter of fact, this is just the equation of harmonic map\cite{23}.

\section{Conclusion}

In summary, we have studied the topological property of the topological
defects in general case by making use of the $\phi $--mapping topological
current theory. As a result, the topological defects are generated from the
zero point of $\phi $--mapping and the topological quantum number of these
defects are the Winding number which are determined by the Hopf indices and
the Brouwer degrees of $\phi $--mapping. The action and the evolution
equations of these defects in Riemannian manifold $G$ are also obtained from
our theory. We would like to point out that the theory of topological
defects in this paper is an unified theory of describing the topological
properties of the arbitrary dimensional topological defects and all the
results are gained from the viewpoint of topology without any particular
models or hypothesis. It is much more important in understanding the origin
and the formation of these topological defects in early universe. Moreover,
as we see in the present work, $\phi $--mapping is a useful method which can
provide an important window into the topological structure of physical
systems.

\end{document}